\documentclass[useAMS,usenatbib]{mn2e}

\title[Proper motion and apparent contraction in the compact radio source J0650+6001]{Proper motion and apparent contraction in J0650+6001}
\author[M. Orienti \& D. Dallacasa]
  {M. Orienti$^{1,2,3}$\thanks{E-mail: orienti@ira.inaf.it},
D. Dallacasa$^{1,2}$\\
$^{1}$Dipartimento di Astronomia, Universit\`a di Bologna, via Ranzani 1,
I-40127, Bologna, Italy \\
$^{2}$Istituto di Radioastronomia - INAF, Via P. Gobetti 101, I-40129
Bologna, Italy\\
$^3$Instituto de Astrof\'{i}sica de Canarias, c/ V\'{i}a L\'actea s/n,
E-38205  La Laguna (Tenerife), Spain\\}

\date{Received \today; accepted ?}

\pagerange{\pageref{firstpage}--\pageref{lastpage}} \pubyear{2002}

\def\LaTeX{L\kern-.36em\raise.3ex\hbox{a}\kern-.15em
    T\kern-.1667em\lower.7ex\hbox{E}\kern-.125emX}

\begin{document}

\label{firstpage}

\maketitle

\begin{abstract}
We present a multi-epoch and multi-frequency VLBI study of the compact
radio source J0650+6001. In VLBI images the source is resolved into
three components. The central component shows a flat spectrum,
suggesting the presence of the core, while the two outer regions, with a
steeper spectral index, display a highly asymmetric flux density. The
time baseline of the observations considered to derive the source
expansion covers about 15 years. During this time interval, the
distance between the two outer components has increased by 
0.28$\pm$0.13 mas,
that corresponds to an apparent separation velocity of
$0.39c \pm 0.18c$ and a kinematic age of 360$\pm$170 years. On the
other hand, a multi-epoch monitoring of the separation between the
central and the southern components points
out an apparent contraction of about $0.29 \pm 0.02$ mas,
corresponding to an apparent contraction velocity
of $0.37c \pm 0.02c$. Assuming that the radio structure is
intrinsically symmetric, the high flux density ratio between the outer
components 
can be explained in terms of Doppler beaming effects where the mildly
relativistic jets are separating with an intrinsic velocity 
of $0.43c \pm 0.04c$ at an angle between 12$^{\circ}$ and
28$^{\circ}$ to the line of sight. In this context, the apparent
contraction may be interpreted as a knot in the jet that is moving
towards the southern component with an intrinsic velocity 
of $0.66c \pm 0.03c$, and its flux density is boosted by a Doppler
factor of 2.0.
\end{abstract}

\begin{keywords}
radio continuum: general - radiation mechanisms: non-thermal -
quasars: individual: J0650+6001
\end{keywords}

\section{Introduction}

The evolutionary stage of a powerful radio source originated by an
active galactic nucleus (AGN) is related to its linear size. In this
context, compact symmetric objects (CSO), which are powerful ($L_{\rm
1.4\,GHz} > 10^{24}$ W/Hz) and intrinsically small (i.e. linear size
LS $<$ 1 kpc) radio sources, should represent a young stage in the
individual radio source life. These objects usually have a roughly
symmetric structure, with core, jets, and mini-lobes/hotspots,
resembling a scaled-down version of the large classical radio galaxies
which conversely have linear sizes of hundreds of kpc or even a few
Mpc. The main characteristic displayed by young radio sources is the
convex synchrotron radio spectrum that peaks at frequencies in the
GHz regime. The spectral turnover is usually due to synchrotron
self-absorption \citep[SSA;][]{mo08,snellen00}, although an additional
contribution from free-free absorption (FFA) has been found in the
most compact objects \citep[e.g.][]{dd09,mutoh02,kameno00}.\\
Conclusive evidence of the genuine
{\it youth} of this class of objects
came from the determination of both kinematic and radiative ages.\\
The determination of the kinematic age is based on multi-epoch
milliarcsecond-scale resolution observations spanning several years,
and able to measure at which rate the hotspots are increasing their
separation. Assuming that the separation velocity $v_{\rm sep}$ has
maintained constant, the kinematic age $t_{\rm kin}$ can be
estimated:\\

\begin{displaymath}
t_{\rm kin} = {\rm LS} \cdot v_{\rm sep}^{-1} \, .
\end{displaymath}

\noindent From the analysis of a dozen of the most compact CSOs
\citep[LS $<$ 20 pc,][]{polatidis03}, it has been derived that the
separation speed is generally in the range of 0.1$c$ and 0.4$c$,
leading to kinematic
ages of a few thousand years.\\
The radiative age can be estimated by multi-frequency observations
able to constrain the radio spectral shape and thus to determine at
which frequency the spectral break occurs. The break frequency
$\nu_{\rm br}$ is strictly related to the radiative lifetime of the
synchrotron emitting electrons $t_{\rm syn}$. Once the magnetic field
is known, for example assuming minimum total energy content
corresponding to equipartition between
particles and magnetic field ($H$) energies \citep{pacho70}, and following some
assumptions \citep{mm03}, the radiative age can be easily estimated by
means of:\\

\begin{displaymath}
t_{\rm syn} = {\rm const} \cdot  \nu_{\rm br}^{1/2} H^{-3/2} \, .
\end{displaymath}

\noindent Several studies aimed at estimating the radiative age in
compact radio objects indicate ages between 10$^{3}$ - 10$^{4}$ years
\citep{mm99,mm03}, in excellent agreement with the kinematic
ages.\\
Given all the assumptions mentioned earlier, these methods do not
provide the accurate source age, that can
be improved either by increasing the time interval spanned by the
observations or through a better sampling of the 
frequency coverage used to derive
the radio spectrum.\\
Among all the CSOs studied in this framework, the radio source
J0650+6001 
represents a peculiar case. This source is
identified with a quasar at redshift $z = 0.455$, and its
optical spectrum is characterized by very weak broad lines and
prominent narrow lines \citep{stickel93}.
It has a non-aligned
triple radio structure with a total angular size of 7 mas ($\sim$40 pc),
and its total radio spectrum turns over around 5.5
GHz \citep{mo07}. 
From a previous study no evidence of proper motion has been found
\citep{mo08}, although the resolution at 5 GHz (i.e. the frequency at
which several observations spanning almost a decade have been
performed) was not adequate to reliably estimate small changes in the
component positions. However, another analysis of the position of the
source components based on global VLBI observations \citep{akujor96},
indicated an apparent contraction between the central and the southern
components. \\
In this paper we report on the results on multi-epoch VLBI and 5-GHz VLA
data of the compact symmetric object J0650+6001, and we present an
interpretation to explain the radio properties of this
source.\\

Throughout this paper, we assume the following cosmology: $H_{0} =
71\, {\rm km\, s^{-1}\, Mpc^{-1}}$, $\Omega_{\rm M} = 0.27$, and
$\Omega_{\Lambda} = 0.73$, in a flat Universe. At the redshift of the
target 1$^{''} = 5.773$ kpc. The spectral index
is defined as 
$S {\rm (\nu)} \propto \nu^{- \alpha}$.\\

\begin{table}
\caption{Log of the archival VLBA observations analyzed in this
paper.}
\begin{center}
\begin{tabular}{ccccc}
\hline
Freq.&Obs. date&Beam&$uv_{\rm max}$&Antennas\\
GHz& &mas$^{2}$&M$\lambda$& \\
\hline
&&&&\\
5.0&22 Nov 1999&2.46$\times$1.94&125&VLBA -Sc -Nl\\
5.0&17 Dec 2004&1.65$\times$0.88&172&VLBA + Eff\\
8.4&17 Dec 2004&0.90$\times$0.54&290&VLBA + Eff\\
&&&&\\
\hline
\end{tabular}
\label{vlba}
\end{center}
\end{table}

\begin{table}
\caption{Multi-epoch VLA flux density at 5 GHz of the source
J0650+6001. Column 1: flux density in mJy; Col. 2: observing date;
Col. 3: Reference: 1: \citet{ulvestad81}; 2: \citet{perley82}; 3:
\citet{odea90}; 4: this work; 5: \citet{dd00}, 6: \citet{tinti05}; 7:
\citet{mo07}; 8: \citet{mo08}.}
\begin{center}
\begin{tabular}{ccc}
\hline
Flux&Obs. date&Ref.\\
\hline
&&\\
906&2 Feb 1979&1\\
830& 18 Nov 1980&2\\
916&30 Dec 1984&3\\
940&21 Jul 1991&4\\
969&23 Feb 1993&4\\
1058&24 Sep 1995&4\\
1236&14 Nov 1998&5\\
1136&3 Jul 2002&6\\
1150&14 Sep 2003&7\\
1106&28 Jan 2004&7\\
1180&19 Nov 2006&8\\
&&\\
\hline
\end{tabular}
\label{oss_vla}
\end{center}
\end{table}

\section{Radio data}

\begin{figure}
\begin{center}
\includegraphics{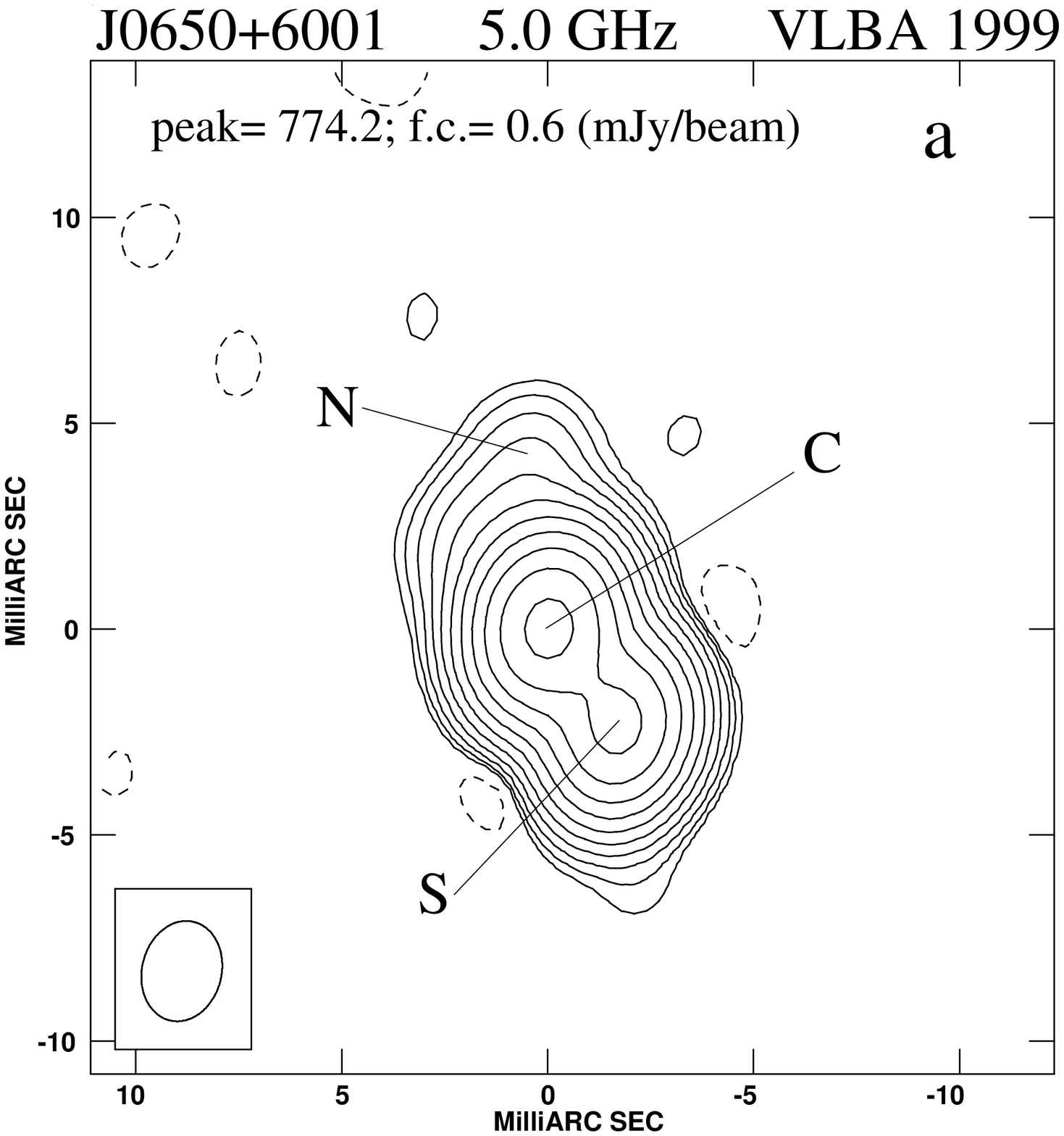}
\includegraphics{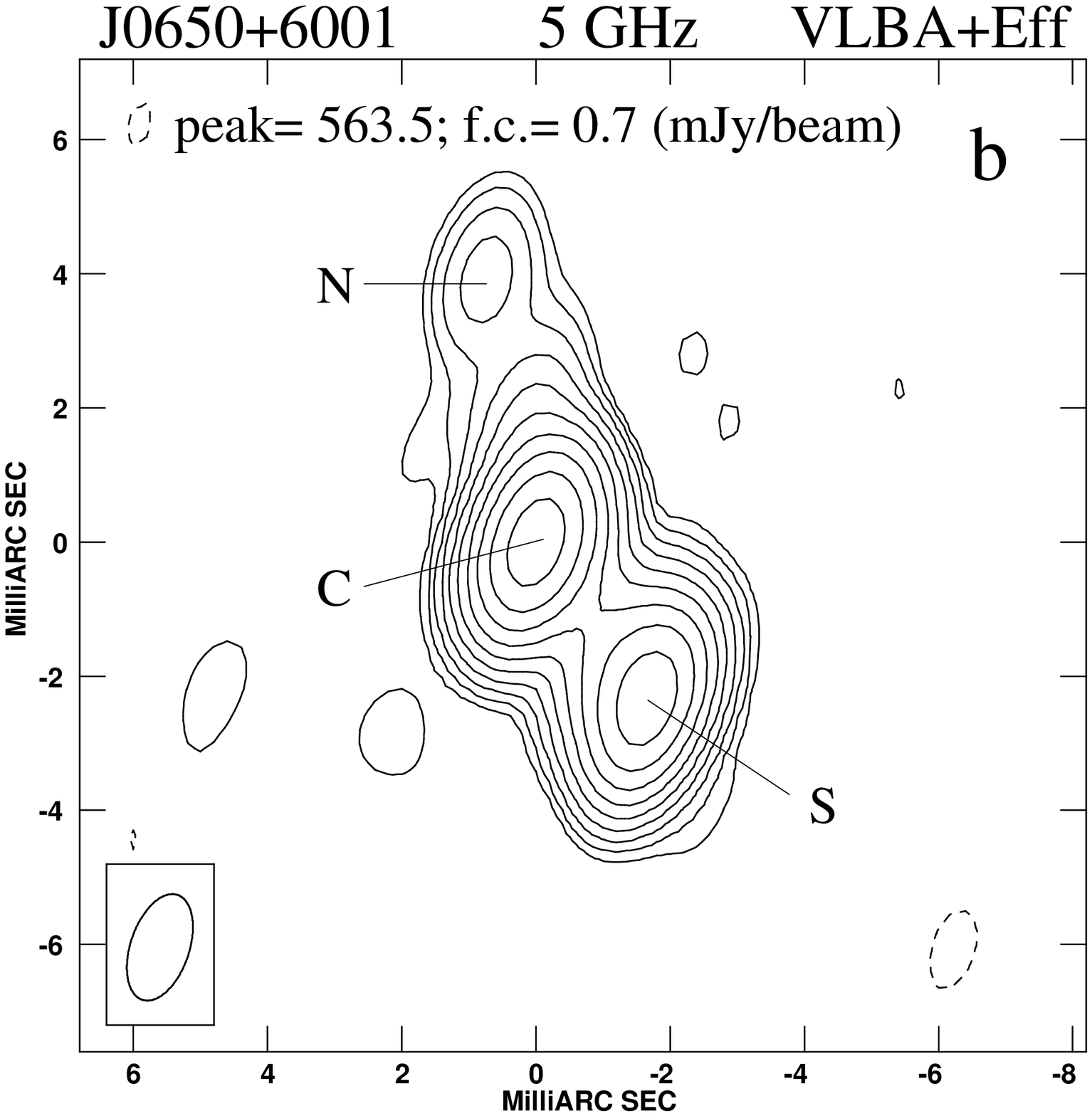}
\includegraphics{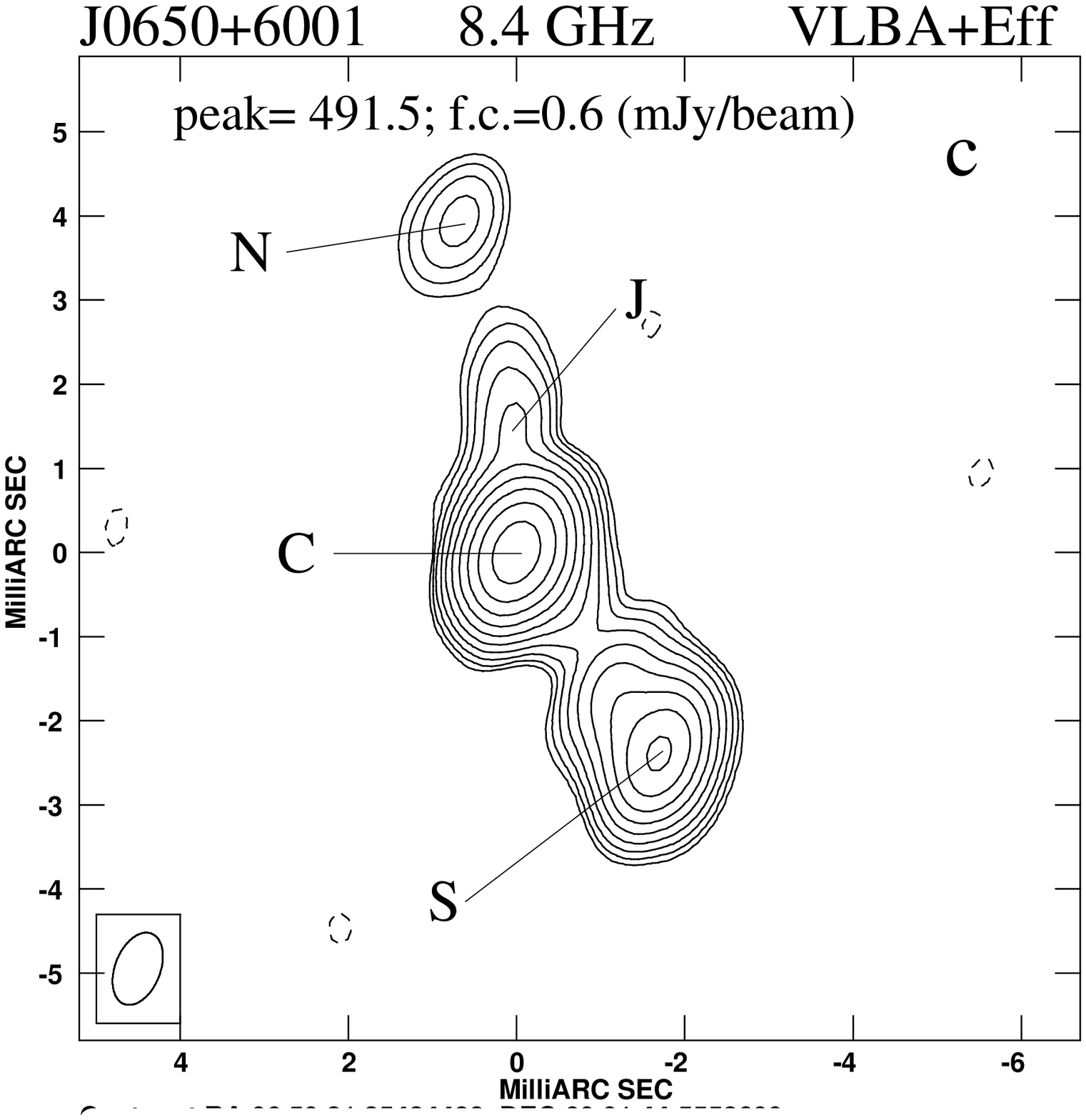}
\vspace{21.5cm}
\caption{VLBA images at 5 GHz (top and centre panels), and at 8.4 GHz
  (bottom panel) of the target source J0650+6001.}
\label{vlba_images}
\end{center}
\end{figure}

To determine a possible source growth, we complemented the information
available in the literature for the target source J0650+6001 with
archival VLBA data obtained in November 1999 at 5.0 GHz, and
VLBA+Effelsberg observations at 5.0 and 8.4 GHz carried out in
December 2004 (see Table \ref{vlba}). The data reduction was carried
out by means of the NRAO AIPS package. The accuracy of the amplitude
calibration has been estimated to be within 5\%. The final images were
produced after a number of phase self-calibration
iterations. Amplitude self-calibration was applied only once at the
end of the process using a solution interval of 30 min, i.e. longer
than the scan length, to remove residual systematic errors and to fine
tune the flux density scale, but not to force the individual data
points to follow the model. \\
To construct a well-sampled light-curve at 5 GHz, we complemented the
information already available in the literature with archival VLA
observations (see Table \ref{oss_vla}). We retrieved and analyzed only
datasets in which the source had been observed for at least 5
minutes in the C band. Since the peak frequency is close to 5 GHz,
observations carried out at frequencies in the range of about 4.9 and 5.0
GHz do not cause significant
scatter on the determination of the light curve.
Data were
reduced following the standard procedure implemented in the NRAO AIPS
package. Uncertainties in the determination of the flux density are
dominated by amplitude calibration errors which have been found to be
within 3\% in all the datasets considered here. VLA images were
obtained after several phase-only self-calibration iterations.\\

\section{Result} 

\subsection{Radio images}

\begin{figure}
\begin{center}
\includegraphics{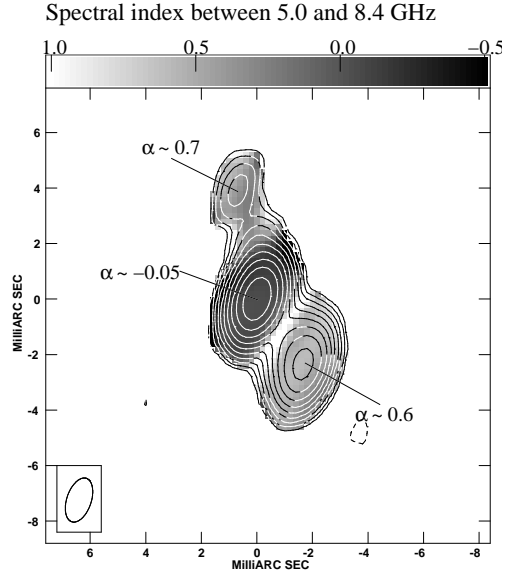}
\vspace{8cm}
\caption{Spectral index distribution across the target source
  J0650+6001 superimposed on the 8.4 GHz contours convolved to the 5
  GHz beam.}
\label{spix}
\end{center}
\end{figure}

Full resolution VLBI images are presented in Fig. \ref{vlba_images}. At
8.4 GHz, in addition to the full resolution image, we also produced a {\it low
resolution} image using the same {\it uv}-range, image sampling and
restoring beam of the image at 5.0 GHz, in order to produce a spectral
index map of the source (Fig. \ref{spix}). No VLA image is presented,
since the source is unresolved on the angular scales typical of this
interferometer. \\
For each VLBI image we provide the observing frequency, the restoring beam
plotted on the bottom left corner, the peak flux density in mJy/beam,
and the first contour intensity ({\it f.c.}) in mJy/beam, that
corresponds to 3 times the off-source noise level. Contour levels
increase by a factor 2.\\
Flux density and angular size of each source sub-component have been
measured by means of the task JMFIT, which performs a 2D-Gaussian fit to
the source components on the image plane. TVSTAT, which performs an
aperture integration on a selected region on the image plane, has been
used to measure the total flux density of the source, which results in
agreement with the sum of the flux densities of the various
sub-components. The parameters derived from the fitting procedure 
are reported in Table
\ref{flux_vlba}.\\

\subsection{Morphology}

When imaged with the high spatial resolution provided by
VLBA+Effelsberg observations, the radio source J0650+6001 shows a
non-aligned triple structure, where the two main components, labelled
C and S in Fig. \ref{vlba_images}, represent almost the 99\% of the total
source radio emission, in agreement with previous observations
\citep{xu95,akujor96,stanghellini99,mo08}. 
Component C, accounting for 63 and 72 mJy at 5 and 8.4 GHz
respectively, possesses a flat spectrum ($\alpha_{5.0}^{8.4} = -0.05
\pm 0.15$), suggesting therefore the presence of the source core. At
8.4 GHz it is clearly resolved in the
North direction, 
indicating the presence of a jet-like component 
(labelled J in Fig. \ref{vlba_images}c) that accounts for 13 mJy.
Component S is located at about 16.2 pc and in p.a. -146$^{\circ}$
with respect to component C, and accounts for 363 and 269 mJy at 5 and
8.4 GHz, respectively. Component N is located at about 20 pc and in
p.a. 13$^{\circ}$ with respect to component C, and accounts for 13 and
9 mJy at 5 and 8.4 GHz respectively. Both components S and N have a
steep spectrum with $\alpha_{5.0}^{8.4} = 0.60 \pm 0.15$ and
$\alpha_{5.0}^{8.4} = 0.70 \pm 0.30$ respectively. 
Furthermore, with a flux
density ratio $S_{\rm S}/S_{\rm N} =$  28 - 30 , depending on the
frequency, component N and S are very asymmetric also in their
observed luminosity.\\

\begin{table}
\caption{Angular separation of components N and C with respect to S
  measured on images at
  5 GHz. Column 1: angular separation between components C and S;
  Col. 2: position angle of component C with respect to component S;
  Col. 3: angular separation between component N and S; Col. 4:
  position angle of component N with respect to component S; Col. 5:
  observing date; Col. 6: reference: 1: \citet{xu95}; 2:
  \citet{britzen07}; 3: this work; 4: \citet{mo08}.} 
\begin{center}
\begin{tabular}{cccccc}
\hline
$d_{\rm C}$&$\phi_{\rm C}$&$d_{\rm N}$&$\phi_{\rm N}$&Observing
date&Ref.\\
mas&deg&mas&deg& & \\
\hline
&&&&&\\
3.03&35.2&6.29&23.4&Jun 1991&1\\
3.08&36.4& - & - &17 Sep 1994&2\\
2.96&36.1&-&-&9 Feb 1998&2\\
2.90&35.9& - & - &22 Nov 1999&3\\
2.81&35.3&6.56&20.6&17 Dec 2004&3\\
2.79&35.6&-&-&28 Jul 2006&4\\
&&&&&\\
\hline
\end{tabular}
\label{flux_vlba}
\end{center}
\end{table}

\begin{table}
\caption{Observational parameters of the components of
J0650+6001. Column 1: source component; Columns 2,3: flux density at 5 GHz from observations
carried out in 1999 and 2004 respectively; Col. 4: flux density at 8.4
GHz; Col. 5: spectral index between 5 and 8.4 GHz; Cols. 6, 7:
deconvolved major and minor axis, respectively; Col. 8: position angle
of the major axis. Spectral index and component angular size are from
observations carried out in 2004. *: the total flux density of the
northern and central components.}
\begin{center}
\begin{tabular}{cccccccc}
\hline
 &1999&\multicolumn{2}{c}{2004}& & & & \\
Comp.&$S_{\rm 5.0}$&$S_{\rm 5.0}$&$S_{\rm 8.4}$&$\alpha$&$\Theta_{\rm
maj}$&$\Theta_{\rm min}$& PA\\
 &mJy&mJy&mJy& & mas&mas&deg\\
\hline
&&&&&&&\\
N&  - & 13&  9& 0.70&1.07&0.27&28\\
C&830*&644&662&-0.05&0.50&0.32&50\\
S& 364&363&269&0.60&0.70&0.38&40\\
&&&&&&&\\
\hline
\end{tabular}
\label{components}
\end{center}
\end{table}

\subsection{Proper motion} 

\begin{figure}
\begin{center}
\includegraphics{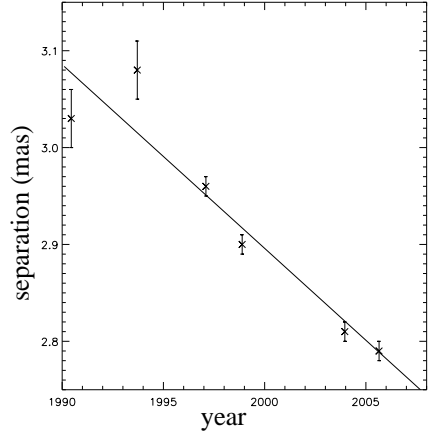}
\vspace{6cm}
\caption{The contraction between the central and the
  southern components of the source J0650+6001. The line represents
  the best fit of $v = -0.02$ mas/year.} 
\label{mas} 
\end{center}
\end{figure}

Decisive evidence supporting the idea that CSOs are radio sources in
an early evolutionary stage has been provided by the measurements of
the proper motion of the hotspots in compact CSOs \citep[see
e.g.][]{polatidis03}. To date, the CSO J0650+6001 lacks a proper measurement
of expansion velocity. On the other hand, an apparent source
contraction has been reported by \citet{akujor96}, representing a
very uncommon case among radio sources.\\
To derive a possible source expansion, we decided to measure the
distance between components S and N from the VLBA+Effelsberg image at
5 GHz. Then we compared the result with what reported in \citet{xu95}
and we found that the separation between these two components has
increased by 0.28$\pm$0.13 mas in 13.5 years (i.e. the time interval
between the two observations). We derive an apparent separation
velocity of $v_{\rm s,a} = 0.39c \pm 0.18c$, that corresponds to a
kinematic age of $t_{\rm kin} = 360 \pm 170$ years. The accuracy of
this separation velocity is limited since only two-epoch observations
are available with adequate resolution and at the same frequency. We
do not consider the VLBA observations carried out in 1999 and 2006
\citep{mo08}, since the resolution was not adequate to reliably
constrain the position of the northern component.\\
To verify the apparent contraction reported by \citet{akujor96}, we
determined the separation between the central and the southern
components from 1999 to 2006, and we compared our results with those
reported by \citet{xu95}, \citet{britzen07}, and \citet{mo08} (see
Table \ref{components}). Then we modelled the data with a linear fit
and we obtain that the separation between components C and S has
decreased of about 0.29$\pm$0.02 mas in 15.6 years (i.e. the time
interval spanned by the observations), corresponding to an apparent
``contraction'' velocity $v_{\rm c,a} = 0.37c \pm 0.02c$
(Fig. \ref{mas}). This suggests that component C likely enshrouds both
the core region and the jet base which cannot be resolved with the
spatial resolution ($\sim$0.6 mas) provided by the observations at
the highest frequency. 
The observed contraction is indicative of a knot in
the jet that is moving towards the southern lobe and it is oriented
with a small angle with respect to the line of sight. For this reason
the flux density of component C is likely dominated by this knot whose
emission is enhanced by boosting effects.

\subsection{Flux density variability}    
 
\begin{figure}
\begin{center}
\includegraphics{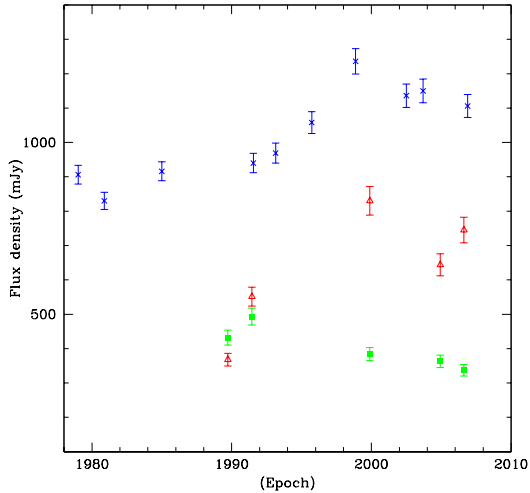}
\vspace{7cm}
\caption{The light curve of J0650+6001 at 5 GHz. Blue crosses indicate the
  source flux density from VLA data (Table 2), while red triangles and green
  squares refer to VLBI flux density of component C and S
  respectively. The error bars refer to 3\% and 5\% of the VLA and
  VLBI flux density respectively (see Section 2). 
  In addition to the VLBI data presented in this paper,
  we include those from Stanghellini et al. (1999), Xu et al. (1995),
  and Orienti \& Dallacasa (2008).}
\label{lightcurve}
\end{center}
\end{figure}

The light curve at 5 GHz of J0650+6001 clearly indicates a steady
increment of the total flux density measured by the VLA over a period
of nearly 18 years (see Section 2). The presence of flux
density variability in a CSO is somewhat surprising, since young radio
sources have been considered the least variable class of radio sources
\citep{odea98}. However, evidence of an increment in the flux density
in the optically-thick part of the spectrum may be easily explained in
terms of source expansion, assuming that the spectral peak is due to
SSA, as in the case of J0650+6001 \citep{mo08}, the source discussed
here. As the source grows, the peak shifts towards lower frequencies,
while the flux density in the optically-thick regime increases
\citep{pacho70}. \\
In Fig. \ref{lightcurve} together with the total VLA flux density, we
report also the VLBA flux density of components C and S. In the light
curve of both the total source and component C, in addition to the 
steady increment of the flux density, it is also present a further
contribution detectable in the 1998 (VLA) and 1999 (VLBA) data. On the other
hand, the light curve of component S is almost constant, with a
possibly slight steady decrease, but without
evidence of significant variability. This behaviour suggests that the
total flux density is dominated by component C, that likely had an
outburst close to the 1998-1999 observing epoch.\\

\section{Discussion}

\begin{figure}
\begin{center}
\includegraphics{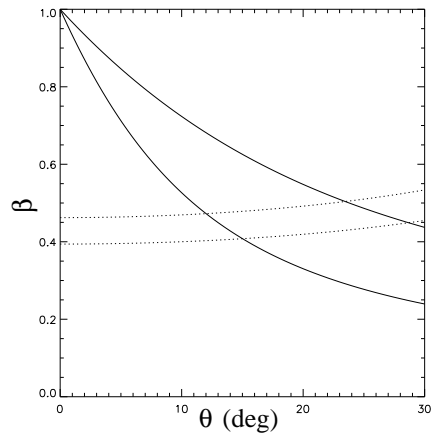}
\vspace{6cm}
\caption{The ($\theta$,$\beta$)-plane for J0650+6001, as derived from
  the northern and southern components. The solid lines represent the
  limits derived from the apparent separation velocity ($\beta_{\rm
    s,a} = 0.39 \pm 0.18$), 
while the dotted lines refer to limits inferred
from the flux density ratio.} 
\label{teta}
\end{center}
\end{figure}

The determination of the properties of the population of young radio
sources is mainly based on the analysis of objects identified with
galaxies, where boosting effects are thought to play a minor
role. Nonetheless, in some of the sources presented in
\citet{polatidis03} also jets have been detected and the proper motion
of knots have been measured (for example in the sources J1945+7055 and
J2355+4950). The target source J0650+6001 is optically identified with
a quasar, and its peculiar characteristics, such as the radio emission
dominated by the core component, the high flux density 
asymmetry between the outer components and their misalignment, may be due to
orientation effects that may modify the intrinsic source
properties. To verify this hypothesis, we consider a beaming model
which assumes that the source is intrinsically symmetric and the
different flux density displayed by components N and S is due to
Doppler boosting. In this context, the flux density ratio is
related to both the jet speed and its orientation $\theta$ with
respect to the line of sight:\\

\begin{equation}
\frac{S_{\rm S}}{S_{\rm N}} = R = \left( \frac{1 + \beta {\rm cos} \theta}{1
- \beta {\rm cos} \theta} \right)^{3+ \alpha}
\label{flusso} 
\end{equation}

\noindent where we have assumed that component S is the one on the
approaching side and $\beta = v/c$, where $c$ is the speed of
light. From VLBI data we found that R = 28 - 30, depending on the
frequency considered, and $\alpha = 0.7 \pm 0.3$. With these values in
Eq. \ref{flusso}, we find the possible combination of $\beta$ and
$\theta$ that is reported in Fig. \ref{teta} between dotted
lines. We must note that although Eq. \ref{flusso} usually applies to jet
components, it can be used also to our purpose, since components N and
S are hotspot dominated and an additional contribution from 
extended features like lobes is marginal since their 
high magnetic field \citep{mo08} causes strong radiative losses at least at the
frequencies considered here. \\
Another way to derive the ($\beta$-$\theta$) combination is from the
measured separation velocity $\beta_{\rm s,a}$ between components N
and S, by means of:\\

\begin{equation}
\beta_{\rm s,a} = \frac{2 \beta_{\rm s,i} {\rm sin} \theta}{1 - \beta_{\rm
s,i}^{2} {\rm cos}^{2} \theta}\; .
\label{beta}
\end{equation}

\noindent Considering $v_{s,a} = 0.39c \pm 0.18c$ 
as determined in Section 3.3, the
resulting possible ($\beta$-$\theta$) combinations are reported in
Fig. \ref{teta} between solid lines. Comparing the areas obtained with
the two different methods, we obtain an intrinsic separation velocity
$v_{\rm s,i} = 0.43c \pm 0.04c$ and a source orientation with respect
to the line of sight of 12$^{\circ}$ and 28$^{\circ}$.\\
The comparison of the flux density of the outer components may be not always
appropriate. In fact, given the relativistic time dilation, we are
watching the components at different stages of their evolution. This
is particularly important in the case of blazars whose jets are
oriented with very small angles $\theta < 5^{\circ}$ to the line of
sight, causing a quite rapid evolution due to 
boosting effects. On the other hand, this effect should not be
relevant for J0650+6001 if we
consider that the lobes/hotspots of this source are oriented at a
relatively larger angle, as also supported by the lack of flux density
variability in the southern component.\\
Another way to determine the  ($\beta$-$\theta$) combination is based
on the comparison of the distance of the two lobes with respect to the
core. However, in this source the central component is likely
dominated by a knot in the jet, thus precluding the determination of
the core position. The apparent contraction between component C and S
is clearly indicating the presence of a compact component moving
towards South with apparent speed $v_{\rm c,a} = 0.37c \pm 0.02c$. If
in Eq. \ref{beta} we consider the apparent separation velocity
computed in Section 3.3 and the same orientation as derived from
Fig. \ref{teta}, we find that this knot has a mildly
relativistic intrinsic velocity $v_{\rm c,i} = 0.66c \pm 0.03c$.\\
With the derived values we can compute the Doppler factor of the jet
knot by means of: \\

\begin{equation}  
D = \frac{1}{\Gamma_{\rm c,i} (1 - \beta_{\rm c,i} {\rm cos} \theta)}
\label{doppler}
\end{equation}

\noindent where $\Gamma_{\rm c,i} = 1 / \sqrt{(1 - \beta^{2}_{\rm c,i})}$
is the Lorentz factor.\\
If in Eq. \ref{doppler} we consider that the knot in the jet in
component C is moving with $v_{\rm c,i} = 0.66c \pm 0.03c$, and it is
oriented with an angle between 12$^{\circ}$ and 28$^{\circ}$ to the
line of sight, we derive a Doppler factor $D \sim 2$, which is consistent
with the peculiarly high observed flux density shown by the central
component. In the case the whole flux density of component C is
attributed to a knot in the jet, we can compute the
intrinsic flux density $S_{\rm intr}$:\\

\begin{equation}   
S_{\rm intr} = S_{\rm obs} D^{-(3 + \alpha)}
\label{boost}
\end{equation}

\noindent where $S_{\rm obs}$ is the observed flux density. If in
Eq. \ref{boost} we consider the observed flux density reported in
Table \ref{flux_vlba}, $D \sim 2$ as derived above, and $\alpha =
0.7$, we find that the intrinsic flux density of component C, 
i.e. core+jet region,
is about 50 mJy at both 5 and 8.4 GHz, 
in good agreement with what expected for the central
component of radio sources.\\
Another explanation for the misaligned source structure and its
high flux density asymmetry may be caused by an inhomogeneous external
environment. 
In fact, in the case the two jets/hotspots experience different
ambient properties on the two sides of the radio source, the side
hitting the most dense medium has its expansion slowed down and its
radio emission enhanced, since the energy losses due to adiabatic
expansion are smaller with respect to the other side. This ends up with
the brighter and smaller radio component lying closer to the core.
If this is the
case, the model developed assuming beaming effects becomes
meaningless, and the source may thus be oriented even at larger angles
to the line of sight, making the estimation of the intrinsic velocity
an upper limit.\\

\section{Conclusions} 

We have presented multi-epoch VLBI and VLA observations of the radio
source J0650+6001. The source shows a peculiar triple structure, where
the outer components are misaligned with respect to the central
flat-spectrum region, and show a high flux density asymmetry. The
comparison between multi-epoch high spatial resolution observations
indicated that the farthest components are increasing their distance
with an apparent separation velocity $v_{\rm s,a} = 0.39c \pm
0.18c$. Assuming that this velocity is representative of the mean
separation speed, we obtain a source kinematic age of 360$\pm$170
years. On the other hand, the distance between the central component
and the southern lobe has decreased with a mean apparent contraction
speed $v_{\rm c,a}= 0.37c \pm 0.02c$. To explain these peculiar
characteristics, we discussed an interpretation based on relativistic
beaming effects. From this analysis we drew a picture in which the source is
oriented with an angle between 12$^{\circ}$ and 28$^{\circ}$ to the line of
sight with the southern component being the approaching side of the
source. The intrinsic separation velocity is $v_{\rm s,i} = 0.43c \pm
0.04c$. The apparent contraction observed between the central and the
southern components may be explained in terms of a knot in the jet that
dominates the radio emission of the central component, that is moving
towards the southern component with a mildly relativistic intrinsic velocity
$v_{\rm c,i} = 0.66c \pm 0.03c$, and with a Doppler factor $D = 2$,
which causes a moderate boost of its flux density. The monitoring of
the flux density variability and the study of the component
separations on a longer time scale will be crucial to better determine
the nature of this source.\\

\section*{Acknowledgments}

We thank the anonymous referee for reading the manuscript and for
giving valuable suggestions.  
The VLA and the VLBA are operated by the US 
National Radio Astronomy Observatory which is a facility of the National
Science Foundation operated under cooperative agreement by Associated
Universities, Inc. This research has made use of the NASA/IPAC 
Extragalactic Database (NED) which is operated by the 
Jet Propulsion Laboratory, California Institute of Technology, 
under contract with the National Aeronautics and Space Administration.

\end{document}